\documentclass [pre,floatfix,superscriptaddress,twocolumn]{revtex4-1}   	

\usepackage{graphicx}				
\usepackage{afterpage}	
\usepackage{amssymb}
\usepackage[version=3]{mhchem}
\usepackage{float}
\usepackage[flushleft]{threeparttable}

\usepackage{color}

\makeatletter
\newcommand{\colorcaption}[2][]{%
	\begingroup%
	\renewcommand{\@caption@fignum@sep}{ (color online). }%
	\caption[#1]{#2}%
	\endgroup%
}

\begin{document}			

\title{Noise-induced distortion of the mean limit cycle of nonlinear oscillators}

\author{Janaki Sheth}
\affiliation{Department of Physics and Astronomy, UCLA, Los Angeles California, 90095-1596, USA}

\author{Dolores Bozovic}
\affiliation{Department of Physics and Astronomy, UCLA, Los Angeles California, 90095-1596, USA}
\affiliation{California NanoSystems Institute, UCLA, Los Angeles California, 90095-1596, USA}

\author{Alex J. Levine}
\affiliation{Department of Physics and Astronomy, UCLA, Los Angeles California, 90095-1596, USA}
\affiliation{Department of Chemistry and Biochemistry, UCLA, Los Angeles California, 90095-1596, USA}
\affiliation{Department of Biomathematics, UCLA, Los Angeles California, 90095-1596, USA}

\begin{abstract}

We study the change in the size and shape of the mean limit cycle of a stochastically driven nonlinear oscillator as a function of noise amplitude. Such dynamics 
occur in a variety of nonequilibrium systems, including the spontaneous oscillations of hair cells of the inner ear.The noise-induced distortion of the limit cycle 
generically leads to its rounding through the elimination of sharp (high curvature) features through a process we call {\em corner-cutting}. We provide a criterion that 
may be used to identify limit cycle regions most susceptible to such noise-induced distortions. By using this criterion, 
one may obtain more meaningful parametric fits of nonlinear dynamical models from noisy experimental data, such as those coming from spontaneously oscillating hair cells. 
\end{abstract}

\maketitle
\section {Introduction}

Nonlinear dynamical models are used to investigate the complex dynamics of many living systems that manifest self-sustained limit cycle oscillations driven by an 
internal energy-consuming process.  Examples include chemical networks underlying the Circadian rhythm, patterns of activity in neuronal networks, 
and cardiac dynamics~\cite{Biological-Clocks1960,Goldbeter1995,Mori1996,Goldbeter2002, Izhikevich2007,Schwab2010}. Another example of such 
an active nonlinear system is provided by the inner ear. The auditory system parses pressure waves, ranging over several orders of magnitude in frequency, 
and detects even {\AA}ngstrom-scale displacements of the mechanically sensitive hair cells ~\cite{Hudspeth2008}. While the mechanisms behind this extraordinary 
sensitivity are not entirely known, previous work suggests that an internal active mechanism amplifies the incoming signal~\cite{Hudspeth14, Robles2001}. The active 
process is also believed to underlie the exceptional sensitivity and frequency selectivity of the auditory system.

Detection of sound in the inner ear is performed by mechano-electrical transducers - bundles of stereocilia - that protrude from the hair cells~\cite{LeMasurier05}. 
The stereocilia contain mechanically gated ion channels that open and close as the bundles are deflected by sound waves~\cite{Vollrath07}. The channels are also 
connected to an internal active motor complex, primarily comprising of Myosin~1c, whose movement along actin filaments regulates the tension in the tip links 
connecting the stereocilia~\cite{Eatock2000}. This interplay of ion-channel gating and myosin motor activity can lead to spontaneous limit cycle oscillations, 
which have been observed {\em in vitro}~\cite{Benser96, Martin2000}.
\begin{figure}
	\includegraphics[width=1\linewidth]{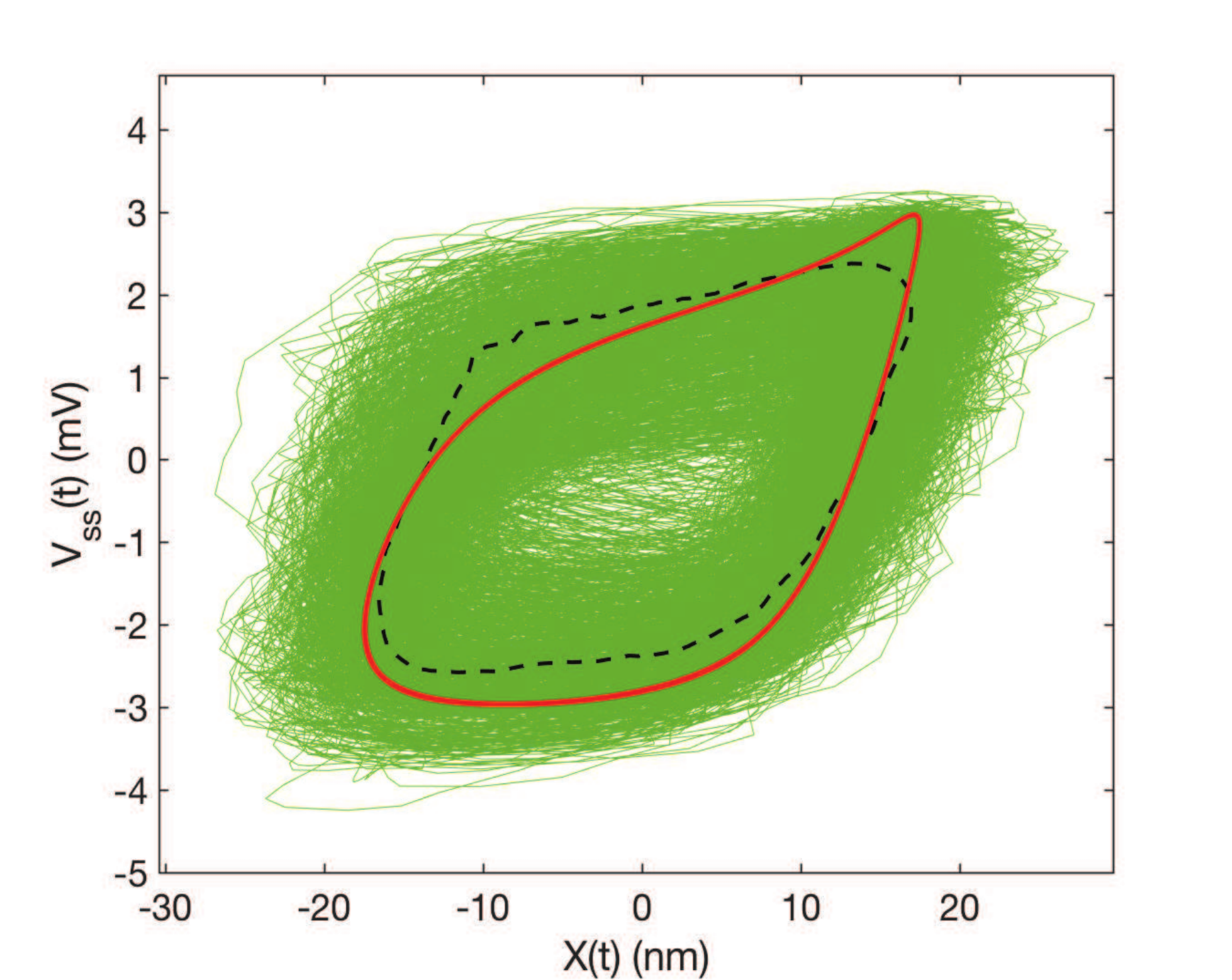}
	\colorcaption{{\bf Stochastic trajectories of the hair bundle model:} A representative stochastic trajectory (green) superimposed upon the deterministic (red) and mean (dashed black) limit cycles.  The noise amplitude is by room temperature and the fluctuation-dissipation theorem $ 2\rm{k_B}T\lambda$, with viscosity $\lambda$. For more details see Ref.~\cite{Nadrowski2004}. 
		\label{fig:corner-cut-hair-bundle}}
\end{figure}

The dynamics of individual hair cells, as well as the overall mechanical response of the inner ear, have been modeled with various systems of nonlinear differential equations of 
multiple levels of complexity~\cite{Reichenbach14, Martignoli2010, Martignoli2013, Gomez2014}. A simple two-dimensional mathematical system exhibiting the 
supercritical Hopf bifurcation, known as the normal form equation, has been shown to reproduce the main aspects of the auditory response~\cite{Camalet2000, Eguiluz2000, Stoop2003}. 
A benefit of simple analytic models is that they account for a number of complex phenomena, such as amplification, compressive nonlinearity, etc., with sparse {\em a priori} 
assumptions, and few free parameters. 

For studies that seek a more direct mapping between variables of the model and underlying physiological processes, more complex models are warranted and have been explored in the literature~\cite{Martin03,Nadrowski2004,Han2010, Yuttana2011}. These models allow for direct comparisons between the numerical predictions and experimentally accessible observables. However, they necessarily include a larger number of fitting parameters and generally have more complex and higher dimensional limit cycles, as they account for more dynamical variables. For example, even a relatively sparse three-dimensional model  that explicitly incorporates stereociliary position, myosin motor activity, and the somatic membrane potential~\cite{Sheth2018} includes many more biologically relevant parameters than the simple two-dimensional models based on the Hopf bifurcation. Given that the experimental records are necessarily stochastic, and typically limited to only a fraction of the total set of dynamical variables in these complex models, the presence of many free parameters in a model raises questions
regarding how to appropriately fit the data. There is an inherent tradeoff between constructing biologically realistic models and limiting the number of free parameters. 

Stochasticity is an inherent feature of hair cell oscillators in particular and biological systems in general~\cite{Netten2003}. Hair bundle motion is affected by thermal Brownian motion from the surrounding fluid; the internal myosin motor complexes are subject to non-equilibrium noise stemming from their attachment and detachment from actin filaments. The membrane potential is affected by ion channel clatter and shot noise in ionic transport~\cite{Nadrowski2004,Dorval2006,Lauger1975}. Hence, even if the macroscopic variables of the system, such as the position of the hair bundle and the somatic membrane potential, obey a low-dimensional dynamical model, these noise sources preclude experimental access to the deterministic limit cycle. Instead, one may observe a number of stochastic but cyclic trajectories and determine a mean limit cycle by averaging over many such trajectories. We previously observed the discrepancy between the deterministic and mean limit cycles, using a three-dimensional model with Gaussian noise. Introduction of stochastic fluctuations not only caused the trajectories to vary from cycle to cycle, but also changed the shape and size of the mean limit cycle. Fig.~\ref{fig:corner-cut-hair-bundle} illustrates differences between the deterministic limit cycle (red) and the average (dashed black) of the hair bundle's stochastic trajectory (green), modeled using noise values corresponding to equilibrium fluctuations at room temperature, as determined by the fluctuation-dissipation theorem~\cite{Sheth2018}. This plot is a mapping of the three-dimensional limit cycle onto the experimentally accessible manifold defined by the bundle deflection and membrane potential measurements~\cite{Meenderink2015}. 

We explore how the mean limit cycle of the stochastic system differs from the deterministic or {\em zero-temperature} limit cycle of the underlying dynamical system. 
By doing so, we 
are able to determine whether noise leads to significant discrepancies between the experimentally accessible dynamics and deterministic theoretical models. Specifically, we explore the 
causes for the rounding of the zero-temperature limit cycle that makes unavailable to experimentalists the sharper features of the deterministic system. 
To explore this question quantitatively, we focus on a generalization of the simple two-dimensional Hopf oscillator, to which we introduce terms that add finer structure to the shape of the deterministic limit cycle. We then are able to observe how these finer details of the shape of the limit cycle are deformed by stochastic forces.

We argue that the generic effect of noise on the limit cycles of dynamical systems is to smooth out the more sharp (high curvature) parts of the trajectory. This effect will impose an upper bound on the useful level of complexity of numerical models, as detailed features, resulting in complex limit cycles in phase space,
will be shown to be experimentally inaccessible.  Our analysis furthermore allows one to determine from the model precisely which features of the limit cycle are most susceptible to noise.  By using that information, one should be able to more meaningfully decide on suitability of various nonlinear models of biological dynamics for interpreting one's data. 

\begin{figure}
	\includegraphics[width=1\linewidth]{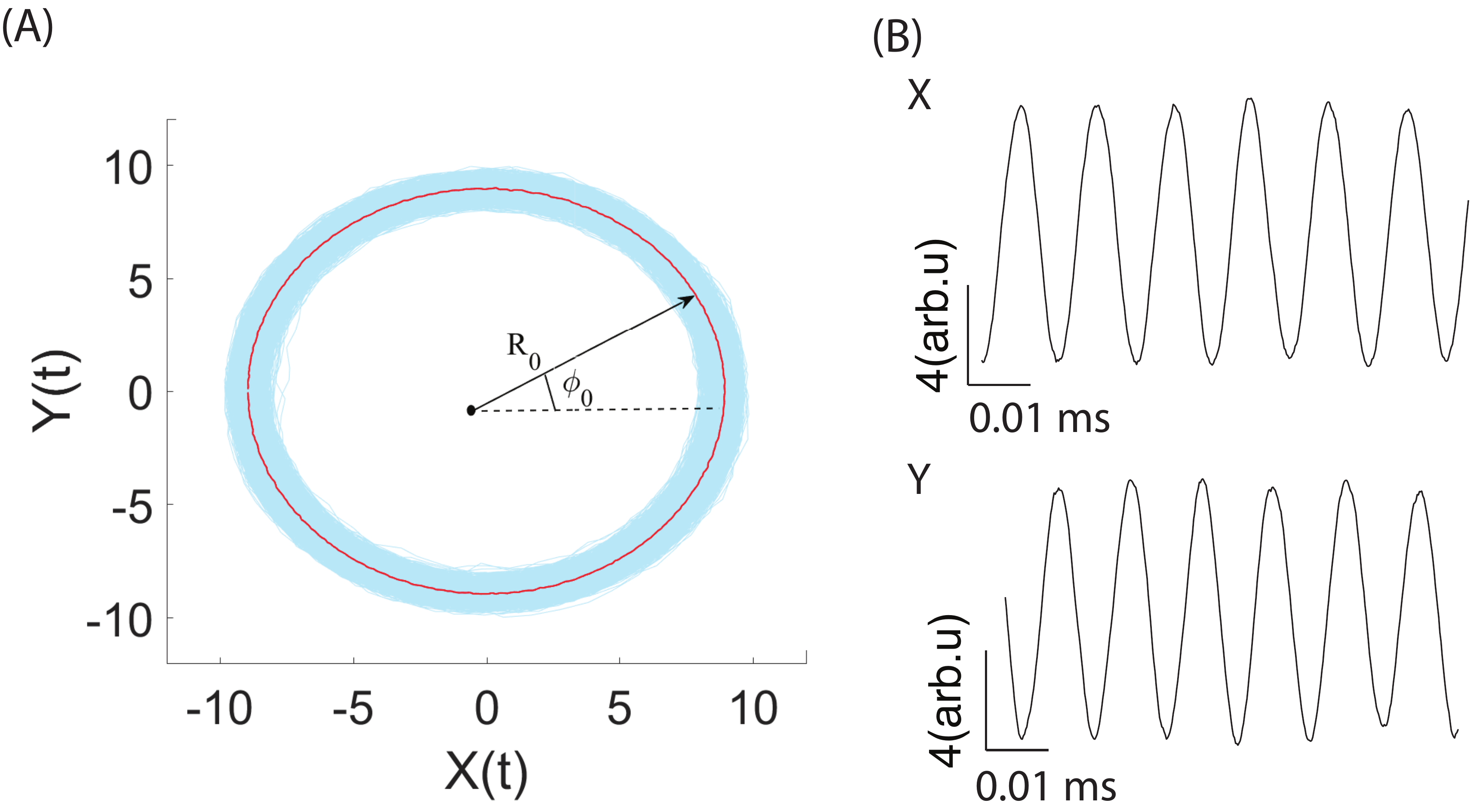}
	\colorcaption{{\bf Numerical simulation of the stochastic Hopf oscillator:} Calculations were performed using Eq.~\ref{Hopf-main}. (A) The finite-temperature (light blue) trajectories and the mean (red) limit cycle. (B) A typical time series (black) of the stochastic dynamics of $X(t)$ and $Y(t)$.  
		\label{fig:Limit-cycle-hopf}}
\end{figure}
The remainder of this article is organized as follows. In section ~\ref{sec:regular-hopf}, we detail a two-dimensional regular Hopf 
oscillator in the stably oscillating regime. In section ~\ref{sec:generalized-hopf}, we analyze the generalized version and illustrate the effects of 
stochasticity and of the internal active drive. Finally, we conclude in section~\ref{sec:summary}, where we review the differences between the experimentally 
accessible trajectory and the theoretical model.

\section{\label{sec:regular-hopf}Regular Hopf oscillator}

\begin{figure}
	\includegraphics[width=1\linewidth]{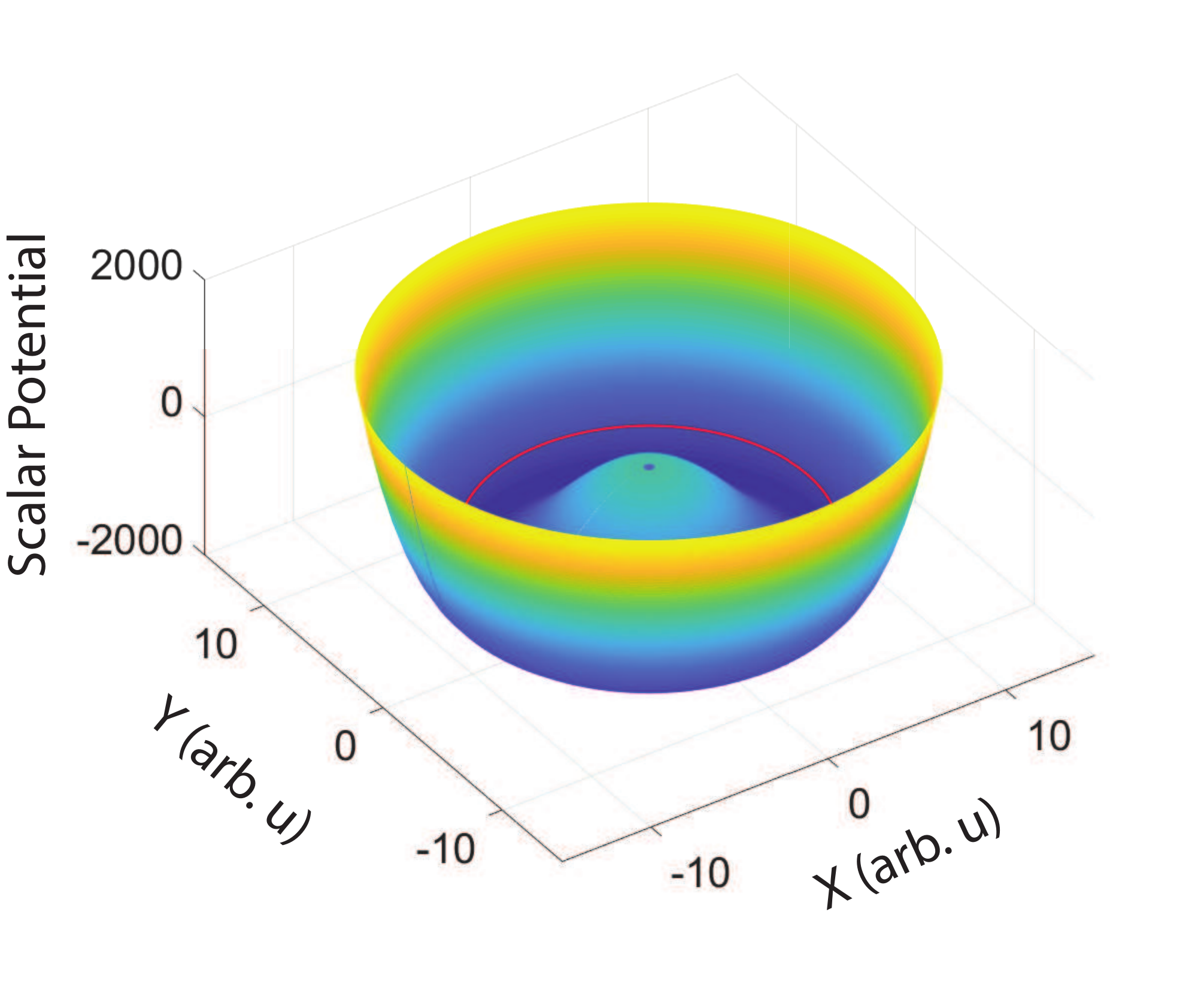}
	\colorcaption{{\bf Hopf Scalar Potential:} The deterministic limit cycle (red curve) lies in the minimum potential region of the 
	Mexican hat potential described by Eq.~\ref{scalar-hopf}. The color map runs from dark blue (low potential) to light yellow (high potential).   The vector potential (not shown) is a constant
	azimuthal vector field which drives the limit cycle dynamics in a counterclockwise circular limit cycle of radius $R_{0}$.  See text for details. 
		\label{fig:Surface-map-regular-hopf}}
\end{figure}
The supercritical Hopf oscillator is the lowest dimensional system ($d=2$) that admits limit cycle oscillations. The normal form of this dynamical system 
can be described in terms of the generalized position variable, $Z(t) = X(t) + iY(t)$, obeying the differential equation
\begin{equation}
\label{Hopf-main}
\dot{Z} = Z \left(  \mu - i \omega \right)  + b Z |Z|^2 + \eta_Z
\end{equation}
The dynamics of the deterministic system depend on the model parameters $\left\{\mu, \omega, b\right\}$. For $\mu>0$, the stable solution is given by the limit cycle of 
radius $R_0 = \sqrt{\mu/b}$ and oscillation frequency $\omega$. 
To fully specify the model, we introduce the stochastic force term $\eta_\alpha$, where $\alpha = X,Y$ are the Cartesian coordinates.  The complex 
noise amplitude discussed in Eq.~\ref{Hopf-main} is related to these two noise terms by $\eta_{Z} = \eta_{X} + i \eta_{Y}$.  Here and throughout this study, we assume that this noise
is uncorrelated, Gaussian white noise with a vanishing mean and the second moment given by
\begin{equation}
\langle \eta_{\alpha}(t) \eta_{\beta}(0) \rangle = 2 T \delta_{\alpha \beta}  \delta(t),
\end{equation}
where $\alpha = X,Y$.
We introduce $T$ as the amplitude of the white noise. We note, however, that in many systems, and in hair cells in particular, the noise may be nonthermal. This 
does not affect our results as long as those nonthermal noise sources are not strongly correlated in time.  Even in that case, we
expect that our qualitative results are not strongly dependent on the assumption of such frequency-independent noise 
amplitudes. However, our results do depend critically on the assumption that the noise amplitude not be too anisotropic.  Strongly anisotropic noise could result 
in a new pattern of noise-induced deformations of the limit cycle distinct from those discussed here.  Similarly, cross correlations between the noise in the $x$ and $y$ channels may 
result in unique stochastic behavior not accounted for here.  

We remind the reader 
that the trajectories of these nonlinear dynamical systems may be thought of as the classical motion of an overdamped particle in $d$ dimensions, moving in response to a force field.  
For a two-dimensional system, the force field may be decomposed into the gradient of a scalar potential, which may be interpreted as the potential energy landscape for the system, 
and the curl of a vector potential.  It is this latter nonconservative force that provides the drive allowing stable limit cycles to exist. The parameters $\left\{\mu, \omega, b\right\}$ may be 
used to define the scalar $(\phi_s)$ and vector $(\phi_v)$ potentials of the system from which one may derive the conservative and nonconservative forces. 
Shortly we will introduce new features into the 
Hopf oscillator model by changing the landscape of effective potential.

The system of dynamical 
equations given by Eq.~\ref{Hopf-main} may be expressed in terms of a two-dimensional vector ${\bf X}(t)  = X (t) \hat{x} + Y(t) \hat{y}$ obeying overdamped motion in a force field ${\bf f}({\bf X})$:
\begin{equation}
\dot{{\bf X}} = {\bf f}( {\bf X} )
\end{equation}
where the force field is given by
\begin{equation}
{\bf f}({\bf X}) = {\bf \nabla } \phi_{s}( {\bf X} ) + \nabla \times {\bf \phi}_{v}({\bf X} ).
\end{equation}

The existence of such a decomposition of the generic vector field ${\bf f}$ is assured by Helmholtz's theorem. For the specific case of the 
Hopf system introduced in Eq.~\ref{Hopf-main}, the scalar and vector 
potentials may be simply computed:
\begin{eqnarray}
\label{scalar-hopf}
\phi_s &=& -\frac{\mu(X^2 + Y^2)}{2} + \frac{b(X^2 + Y^2)^2}{4}\\
\label{vector-hopf}
{\bf \phi}_v &=& -\frac{\omega (X^2 + Y^2)}{2} \hat{z}.
\end{eqnarray}

The scalar potential has one of two forms depending on the sign of $\mu$.  For negative values, the potential has a single minimum at the origin, and the deterministic 
dynamical  system has a single fixed point. For positive $\mu$, the origin is a local maximum of the scalar potential, and a new set of local minima appear on the 
circle of radius $R_{0} = \sqrt{\mu/b}$ about that center.  This form of the potential is the well-known ``Mexican hat'' shown in   Fig.~\ref{fig:Surface-map-regular-hopf} .  For 
finite values of the drive $\omega > 0$, we observe that the curl of the vector potential $ {\bf f}_{v}= \nabla \times \phi_v$ is tangent to the circular ring. It drives the ${\bf x}$
variable anticlockwise along the limit cycle, defined  by the circular ring of minima. The transition between the stable fixed point and the stable limit cycle of angular velocity $\omega$ occurs at $\mu=0$ and is known as the supercritical Hopf bifurcation.  

Turning to the motion of the stochastically driven system, we observe that, across a range of noise amplitudes, the trajectories remain constrained to the trough of the 
scalar potential at $R_{0} = \sqrt{\mu/b}$ that stabilized the deterministic limit cycle.   
Because the scalar potential is locally symmetric for positive and negative radial displacements from the limit cycle and since the 
vector potential has no radial component, the mean limit cycle of the stochastic system is identical to the deterministic one.  Fig.~\ref{fig:Limit-cycle-hopf} illustrates these stochastic dynamics.

In Fig.~\ref{fig:Limit-cycle-hopf}A, we show a representative trajectory (light blue) superposed upon the 
mean limit cycle (red). We plot in Fig.~\ref{fig:Limit-cycle-hopf}B typical $X(t),Y(t)$ traces, as might be 
obtained from hair cell data. Herein, $\mu = 80, b = 1, \omega = 200$, and the details of the simulation are described in Appendix A. The mean limit cycle for the finite-temperature 
system is computed by binning the phase space $\left\{-\pi, \pi\right\}$ into 200 bins and averaging over multiple trajectories. For this simple model of a 
Hopf oscillator, the average cycle is similar to the deterministic limit cycle, due to the high symmetry of the system. When the 
potential landscape of the system is more complex (i.e. exhibits lower symmetry), 
this correspondence between the mean and deterministic limit cycles no longer holds.  We study the lower-symmetry, generalized Hopf system in the next section.

\section{\label{sec:generalized-hopf}Generalized Hopf oscillator}

\subsection{Model and dynamical phase diagram}
To explore the effects of noise on the mean limit cycle, we add symmetry-breaking terms to the Hopf oscillator by changing the scalar potential $\phi_s$. 
\begin{widetext}
\begin{eqnarray}
\label{scalar-general-hopf}
\phi_s &=& -\frac{\mu(X^2 + Y^2)}{2} + \frac{b(X^2 + Y^2)^2}{4} + \alpha cos(n\theta)e^{-(\sqrt{X^2 + Y^2} - \sqrt {\frac{\mu}{b}})^2}\\
\label{vector-general-hopf}
\phi_v &=& -\frac{\omega (X^2 + Y^2)}{2} \hat{z}
\end{eqnarray}
\end{widetext}
The modulation introduces $n$ local maxima (and an equal number of local minima) to the scalar potential that remove the azimuthal symmetry present in Eq.~\ref{scalar-hopf}.  By tuning the radial 
position of those extrema to the center of the circular trough of the Hopf potential, we force trajectories near the previous limit cycle to deform and can control that deformation by the 
strength of the perturbation $\alpha$.  Here, we consider the case of a four-fold potential landscape, $n=4$, but we believe that none of the results shown below depend critically on 
that choice.  
\begin{figure}
	\includegraphics[width=1\linewidth]{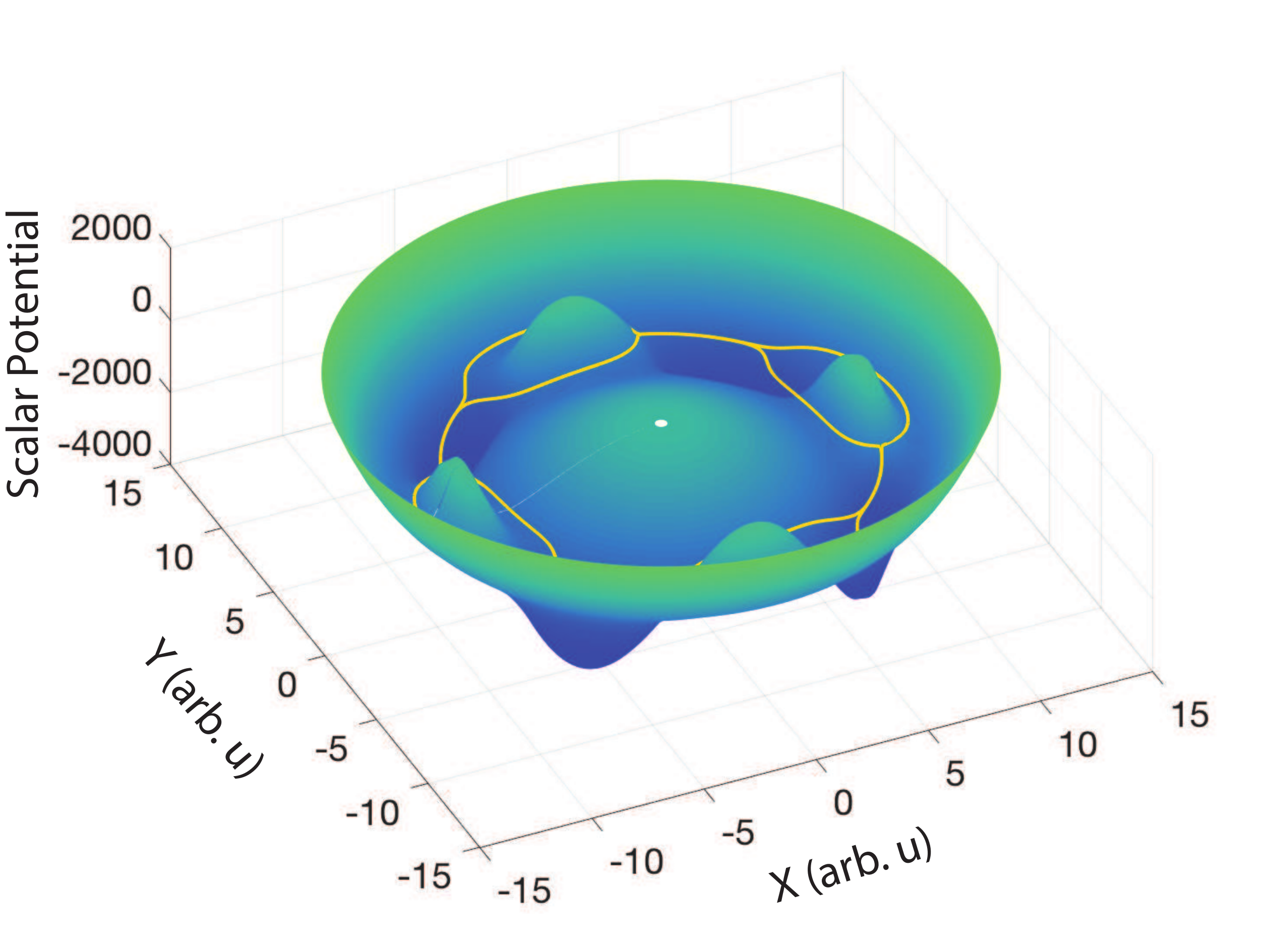}
	\colorcaption{{\bf Scalar Potential map for generalized Hopf:} 3d plot of the scalar potential in Eq.~\ref{scalar-general-hopf}, for $n = 4$ with the valleys seen in dark blue and hills in between 		them. The color map spans across dark blue (low potential) to light green (high potential). The deterministic limit cycle (yellow) for small vector potential skirts 
	around the hills and pinches at the valleys. 
		\label{fig:Surface-map-daisy-hopf}}
\end{figure}

Fig.~\ref{fig:Surface-map-daisy-hopf} shows the modified scalar potential for $n=4$, along with the deterministic limit cycle shown in yellow.   It should be noted that there are pairs of 
degenerate paths about each of the local maxima. These minima also introduce new fixed points that remain stable for sufficiently small values of the vector potential.  To 
find stable limit cycles we require that strength of the vector potential exceed
\begin{equation}
\label{omega-condition}
 \omega^{\star}  = n b \frac{\alpha}{\mu}.
 \end{equation}
Beyond this point, stable limit cycles exist, but their shape continues to change with increasing vector potential strength $\omega$.  We study these dynamics for various values of 
$\omega/\omega^{\star}$.

\begin{figure}
	\includegraphics[width=1\linewidth]{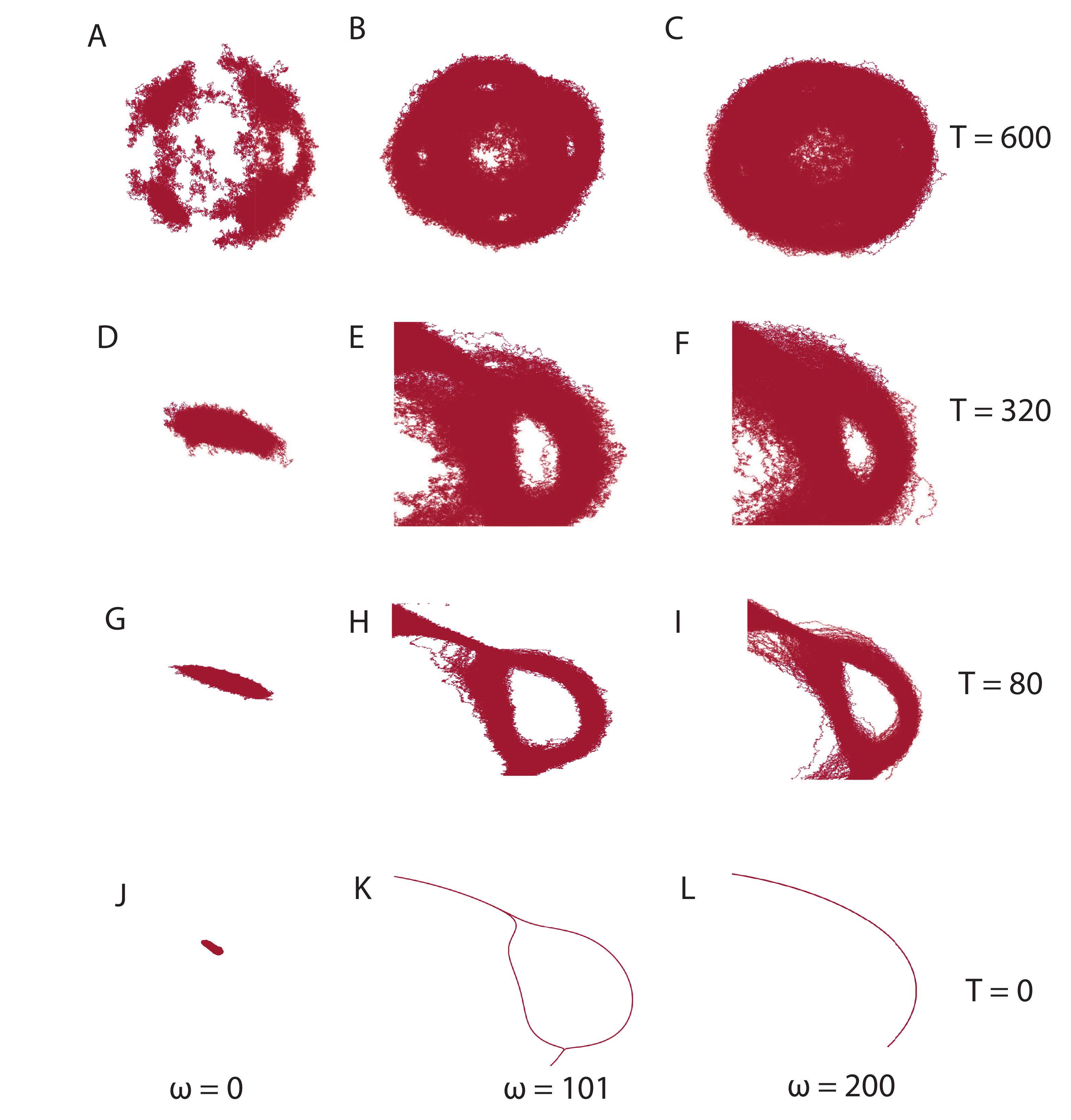}
	\colorcaption{{\bf Stochastic trajectories with variation in temperature and $\omega$:} Quarter lobes of the trajectories obtained by solving Eq.~\ref{scalar-general-hopf} using $\omega$ values of $\left\{0, 101, 200 \right\}$ and $\langle \eta_z^2 \rangle$ values of $\left\{0,80,320,600 \right\}$. 
		\label{fig:Temp-omega-plot}}
\end{figure}

Introduction of the local minima renders the dynamical phase diagram more complex. In Fig.~\ref{fig:Temp-omega-plot}, we show this phase diagram under varying noise amplitude $T$ and drive frequency $\omega$. The top row of the phase diagram shows the full limit cycle, while the lower rows zoom in on one of the four equivalent local quadrants of the system.  The $\omega=0$ column shows the expected behavior of an equilibrium system with increasing levels of noise. For sufficiently small $T$, stochastic trajectories are confined to one of the four local minima (we show one such case in the figure).  The trajectories deviate further from the minimum of the potential with increasing $T$, as one expects in the vicinity of a fixed point. 
Over sufficiently long times, one observes thermally activated hopping between these minima, so that the system diffuses around the ring set by the underlying circularly symmetric potential.  Alternatively, the same behavior can be observed by increasing $T$ at fixed length of the trajectories. The sequence (J, G, D, and A) in Fig.~\ref{fig:Temp-omega-plot} demonstrates these effects.  

In the case of small but finite $\omega$, the drive biases the hops between local minima to favor those in the anticlockwise direction, along the force generated by the vector potential.  However, if one chooses $\omega < \omega^{\star}$, the drive is not sufficient alone to drive transitions between 
local minima, and the deterministic system remains trapped within one of these wells of the scalar potential. 
In this study, we are primarily interested in the case where the deterministic system has a stable limit cycle, so we begin our studies 
where the vector potential is just strong enough to destabilize the local minima. A sequence of such minimally stable deterministic trajectories with increasing noise amplitude can be see in (K, H, E, B) of 
Fig.~\ref{fig:Temp-omega-plot}.  We note that even small values of the noise amplitude are capable of allowing the dynamical system to 
explore both trajectories about the local potential maxima. 

Finally, with a sufficiently strong vector potential (here $ \omega = 200$),  the deterministic system (and the system with sufficiently small noise amplitude) approaches the circular limit cycle of the standard Hopf oscillator with the circularly symmetric diving force overwhelming the symmetry-breaking scalar potential. This is shown in panel L of Fig.~\ref{fig:Temp-omega-plot}. 
Upon increasing the noise amplitude, as shown in the sequence (L, I, F, C) of Fig.~\ref{fig:Temp-omega-plot}, we
observe both paths around the local maximum appearing once again. Since the limit of very large drive restores the circular symmetry, and since we aim 
to study the noise-induced loss of fine detail in more complex limit cycles, the large $\omega$ limit will not be considered further.
In panels H and I, one observes a dispersion of trajectories around the point where the inner path about the 
local maximum reconnects with the outer path.  This localized broadening is an example of noise-activated {\em corner cutting} in the generalized Hopf model that is the focus
of this manuscript.

\subsection{Noise-induced corner cutting}

\begin{figure}
	\includegraphics[width=1\linewidth]{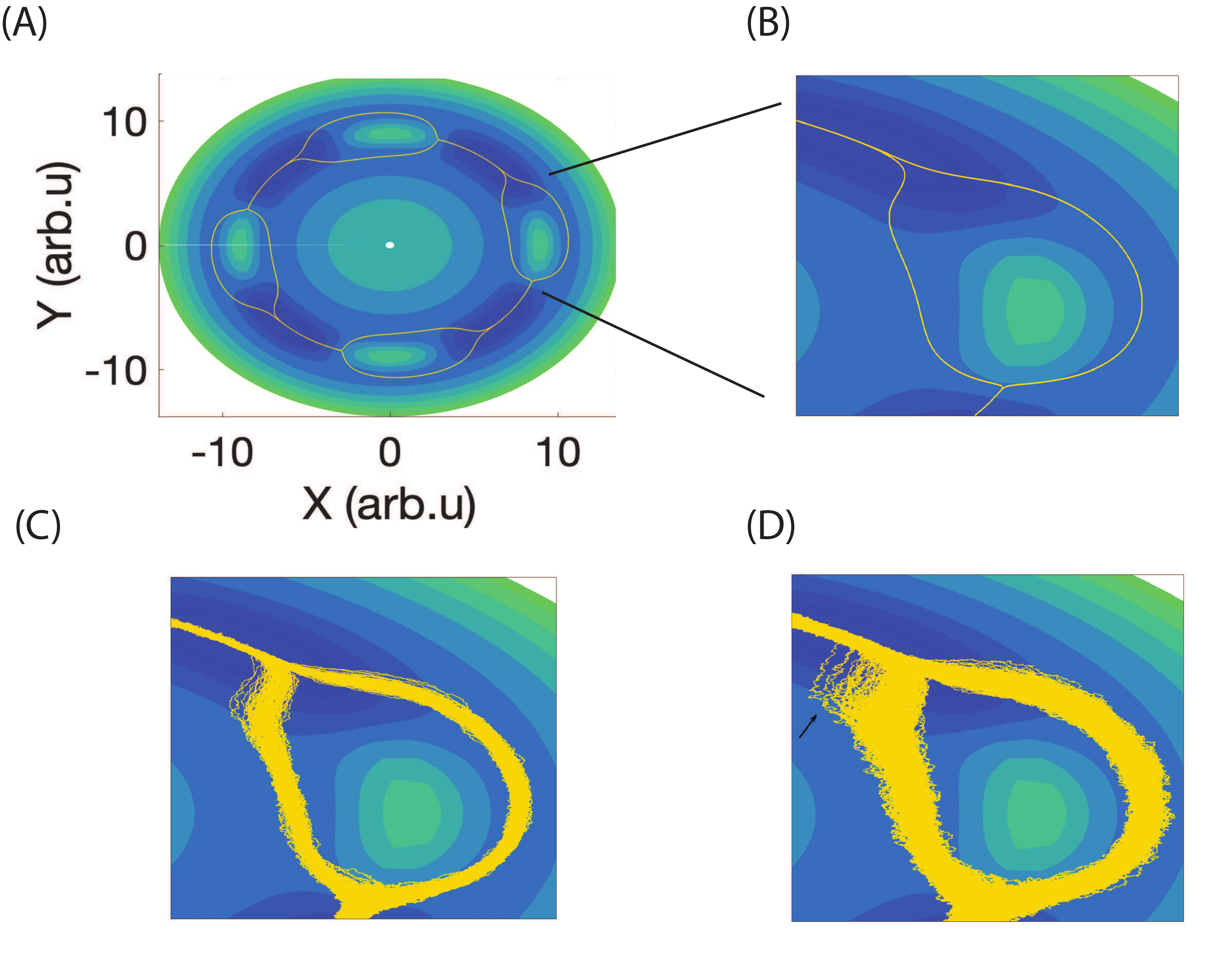}
	\colorcaption{{\bf Examples of corner cutting:} (A) The deterministic oscillator tracks the local minimum potential regions. (B) One lobe of the potential landscape. 
	(C) Oscillator at $\langle \eta^2 \rangle = 10$. (D) Oscillator at $\langle \eta^2 \rangle = 30.$ A (black) arrow points to an example of a corner cutting trajectory. These have been simulated using parameter values $\mu = 80, b = 1, \alpha = 2000, n = 4$, resulting in  $\omega^{\star} = 100$ -- see Eq.~\ref{omega-condition} -- and  $\omega = 101.$
		\label{fig:Noise-daisy-hopf}}
\end{figure}

We now explore in detail the noise-induced corner cutting at intermediate values of both noise amplitude and drive, consistent with panel H in the phase diagram. 
As expected, the deterministic oscillator occupies the low potential regions at nearly all phases of the oscillation (see Figs.~\ref{fig:Noise-daisy-hopf}(A),(B)). However, 
upon increasing the noise amplitude in the system, as shown in panels (C) and (D) of Fig.~\ref{fig:Noise-daisy-hopf}, the trajectories deviate from the $T=0$ curve by cutting across the sharper (higher curvature) features of the deterministic path.  

The net effect of these devations is that the mean shape of the limit cycle increasingly deforms with noise amplitude.  In 
particular, the higher curvature features of the deterministic limit cycle, apparent where the inner path (smaller radius) around the local maximum converges with the outer path, are lost with 
increasing noise amplitude.  We refer to this phenomenon as {\em corner cutting}, since the sharper corners of the deterministic limit cycle are smoothed out. 

The corner cutting observed in the generalized Hopf model resembles that observed in the hair cell oscillator model. One observes in Fig.~\ref{fig:corner-cut-hair-bundle} the noise-induced rounding of the high curvature corner in the upper right quadrant of the deterministic limit cycle. Comparing panels C and D of Fig.~\ref{fig:corner-cut-hair-bundle}, we see that increasing noise amplitude increases both the frequency at which paths deviate from one that 
follows the local potential minimum and the degree of their deviations, indicating that this effect is indeed driven by stochastic processes.  

The degree of corner cutting at different points along the deterministic trajectory, which exhibit the same scalar potential, are not equivalent.  For 
example, we do not observe as much corner cutting at the point where the limit cycle diverges when approaching the local maxima as where these paths converge on the 
other side.  This shows that phenomenon is not simply a feature of the local scalar potential, which is the same at both 
of these points. 
\begin{figure}
	\includegraphics[width=1\linewidth]{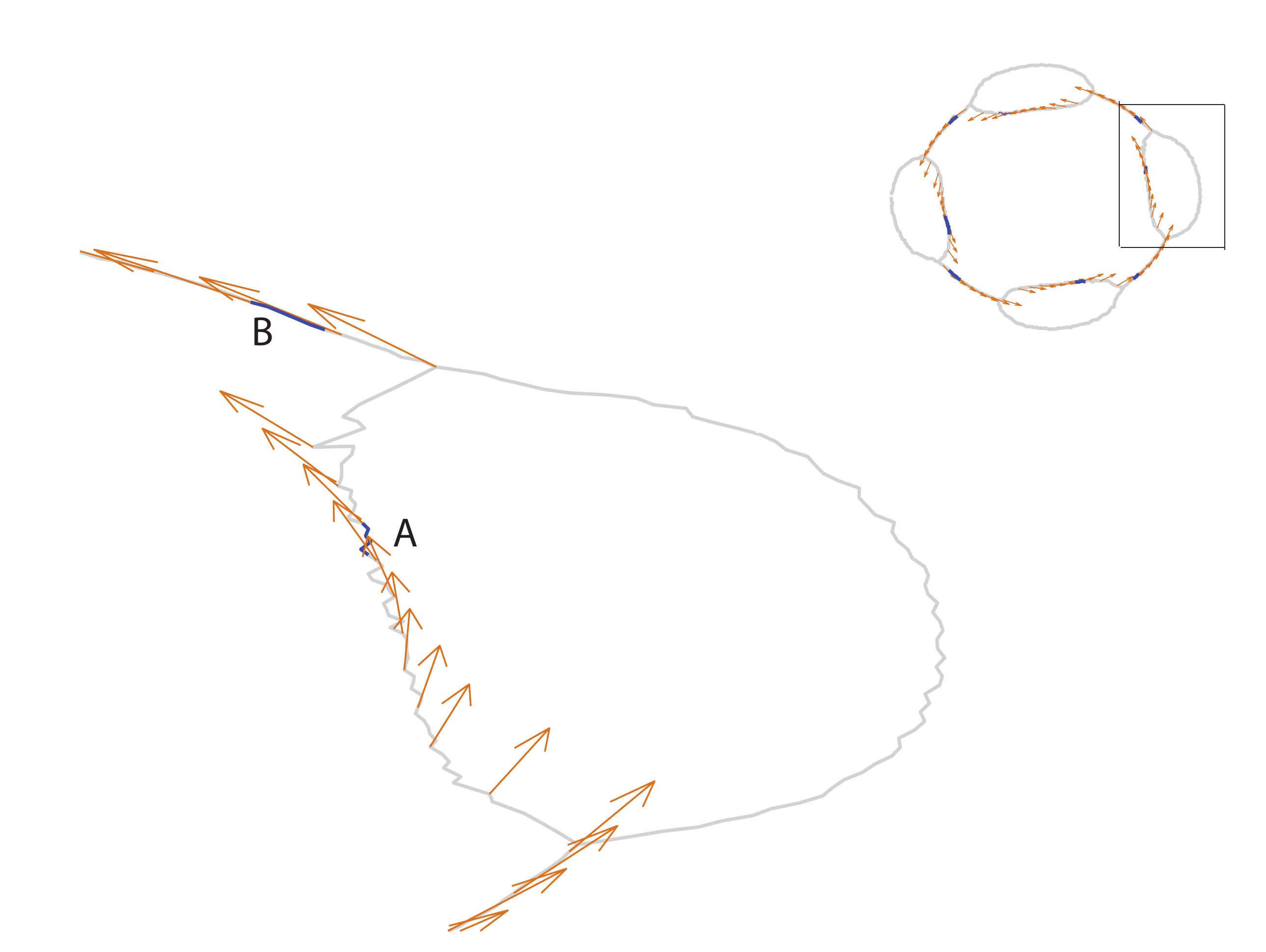}
	\colorcaption{{\bf Direction of $f_v$:} One of the lobes of the mean limit cycle (grey) of the $ \langle \eta^2 \rangle = 30$ stochastic system, with the (blue) regions (A,B) corresponding to arc lengths amidst which the corner cutting trajectories deviate from the particle's average behavior. The direction of ${\bf f}_{V}$ is illustrated by (orange) arrows. This lobe corresponds to the marked lobe in the (upper right) inset.
 		\label{fig:fv-30T}}
\end{figure}

In Fig.~\ref{fig:fv-30T}, we plot the drive force associated with the vector potential ${\bf f}_{V}$ along the mean limit cycle. 
The mean limit cycle is calculated in a similar manner as the regular Hopf oscillator, with an additional calculation at each phase to check for the 
presence of one or two maxima in the trajectory density. The peaks are considered distinct if they are radially separated from $R_{0 = }\sqrt{\mu/b} = \sqrt{80}$ by a 
distance of 0.2 or more. We identify the corner-cutting paths as events that lie at a potential energy greater than $3T$ 
compared to the potential energy of the mean curve.

One immediately observes the distinction between the entry and exit points
of the loop around the local potential maximum. Near the entry point, the vector potential force is tangent to the path of the limit cycle.  Near the point where the inner path merges with the 
outer one, however, the drive force has a significant component normal to the mean path.  The drive force plays a role in enhancing the thermally excited deviations from the 
mean limit cycle.  Moreover, asymmetric deviations from the deterministic limit cycle resulting in deformation of the mean limit 
cycle from the deterministic one are strongest in regions where two conditions are met. First, there must be a significant component of the drive force normal to the deterministic limit cycle, and second, the confining 
potential about that limit cycle must be weak.  

\begin{figure}
	\includegraphics[width=1\linewidth]{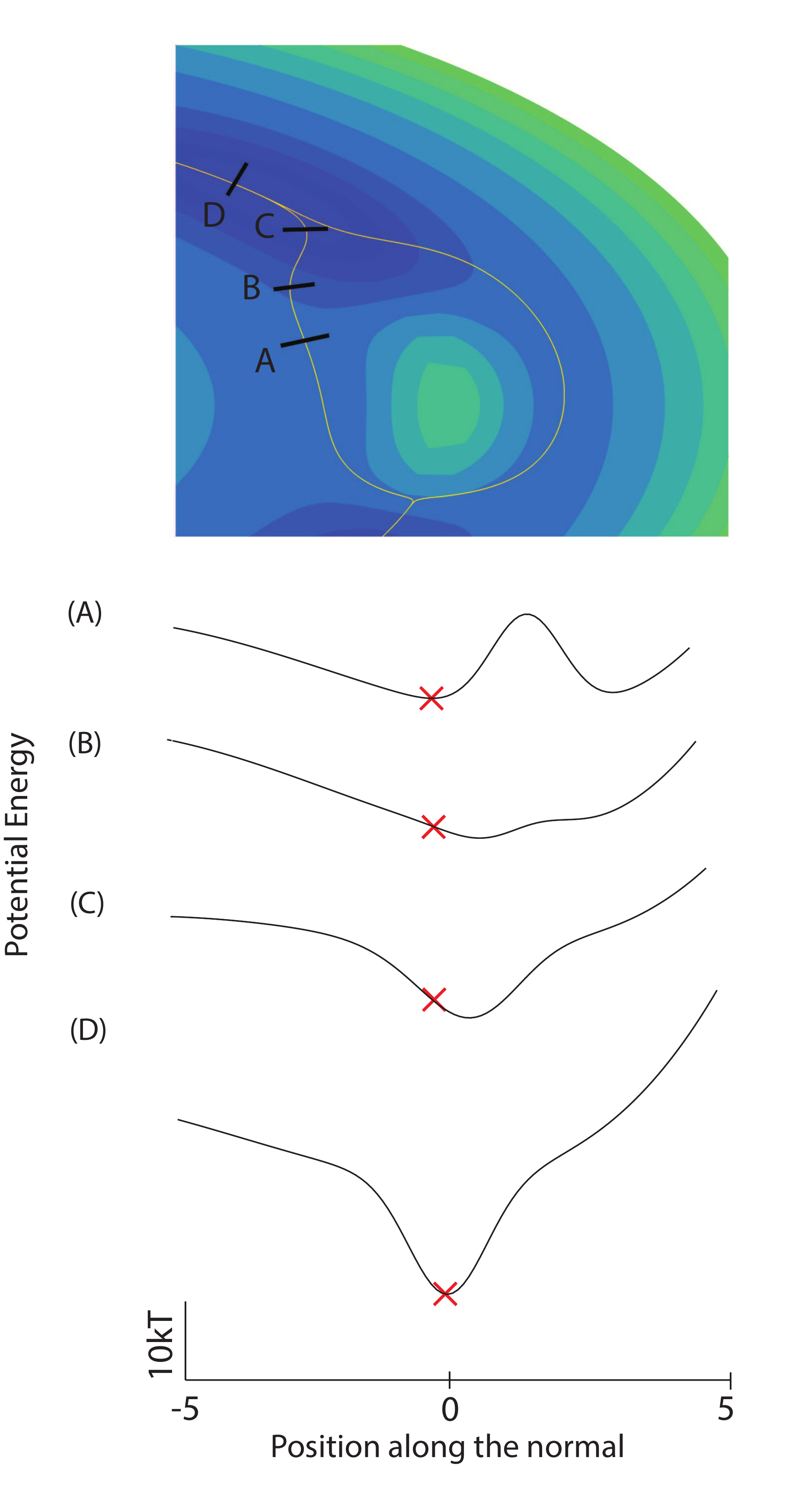}
	\colorcaption{{\bf Confining potential:} (A-D) Energy landscapes in the $\hat{n}$ direction to the zero-temperature limit cycle, corresponding to the A-B arclength in Fig.~\ref{fig:fv-30T}. The (red) cross is indicative of the noiseless particle position, with negative values pointing towards $(0,0)$. These positions correspond to the (black) cuts along the (yellow) limit cycle atop.
		\label{fig:Potential-energy-along-limit-cycle}}
\end{figure}

In the lower panel of Fig.~\ref{fig:Potential-energy-along-limit-cycle}, we plot the potential energy versus distance, measured along the local normal to the mean limit cycle. Plots are obtained at various 
points (labeled A, B, C, D) along that limit cycle, indicated in the upper panel of the same figure.  The potential in the normal cross section at A shows two minima, consistent with the 
two paths of the deterministic limit cycle at that point.  Upon approaching the junction of those two paths at B and C, one sees these two local minima merge into a single broader minimum.
This minimum then deepens, and the confining potential sharpens, as one moves alway from the local maximum (shown at D).  
It is clear that the combination of the weak (small curvature) confining potential and large normal component of the drive force at points near B make this area most susceptible to noise-induced 
trajectories escaping from the mean path.  The asymmetry of those escapes, {\em i.e.}, their preference for moving to smaller radii, leads to an enhancement of the
noise-induced distortion of the mean limit cycle near B.  

\begin{figure}
	\includegraphics[width=1\linewidth]{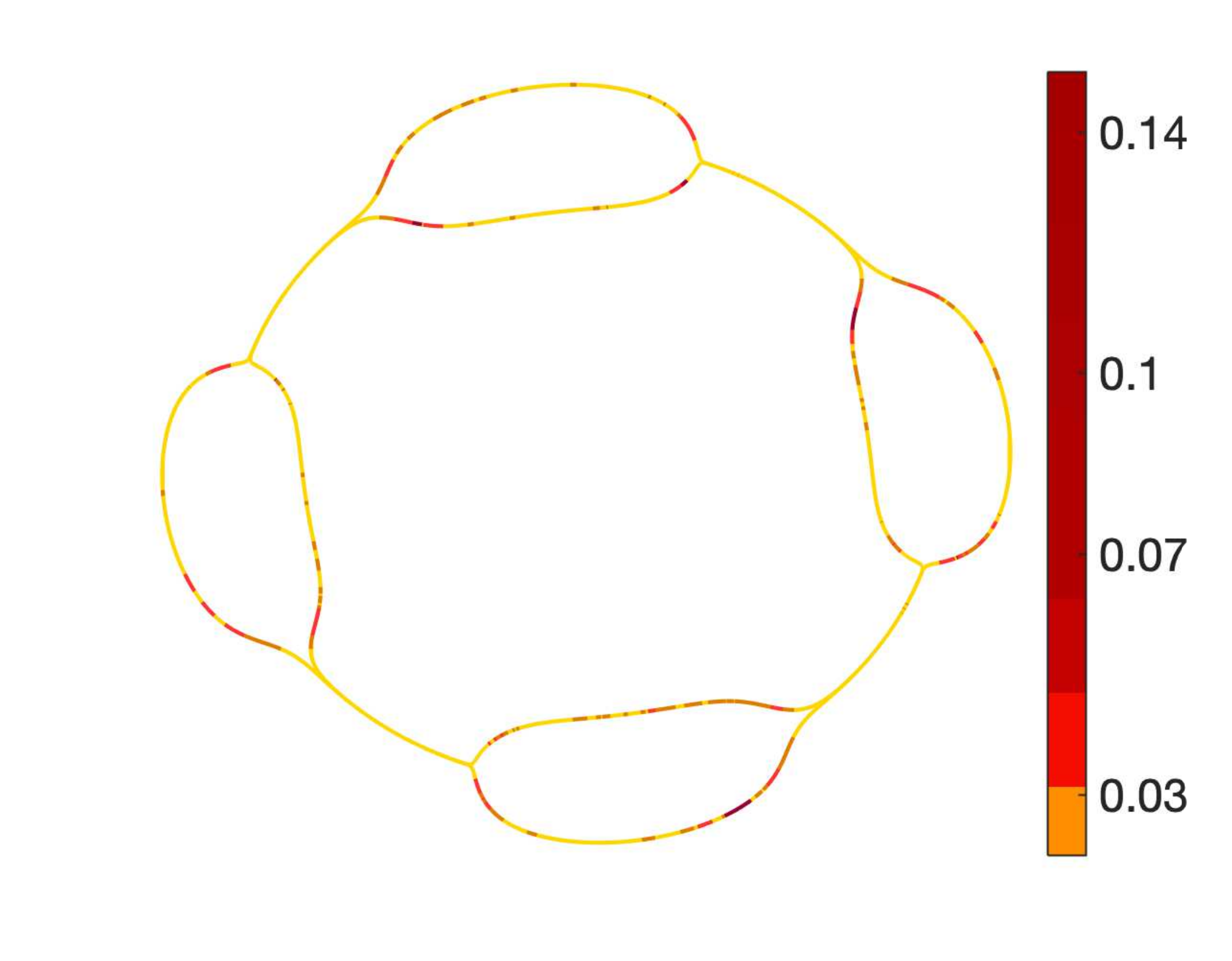}
	\colorcaption{{\bf Nonuniform distortion of the limit cycle:} The distance of the mean limit cycle at $\langle \eta^2\rangle = 2000$ from the underlying noiseless curve shown as a color map. These values are normalized to the average value of the zero-temperature cycle, $\sqrt{\mu/b}$. 
		\label{fig:Heatmap-low-omega}}
\end{figure}

Based on the above analysis we expect that the effect of noise on the mean limit cycle of the oscillator depends strongly on the arclength. In other words, different regions along the 
deterministic limit cycle deform differently with increasing noise amplitude so that the shape of the limit cycle itself changes with noise amplitude. To investigate this effect, we measure 
the normal distance between the deterministic limit cycle and the one measured at ``high temperature'',  where $\langle \eta^{2} \rangle = 2000$.  In Fig.~\ref{fig:Heatmap-low-omega}, 
we color the deterministic limit cycle using a heat map to represent this noise-induced deformation.  In that figure, the cooler (yellower) colors depict smaller 
noise-induced distortions. The deformation is clearly nonuniform along the limit cycle (although still symmetric under
rotations of the figure by $\pi/4$ due to the underlying symmetry of the $n=4$ perturbation).  The greatest deviations occurs at the region 
corresponding to (A-D) of Fig.~\ref{fig:Potential-energy-along-limit-cycle}, showing that the principal cause of these distortion ``hot spots'' are the 
corner-cutting trajectories where the two limit cycle arcs converge at the end 
of the local potential maxima.   
 
\subsection{Predicting regions of noise-induced limit cycle distortion}
To better understand the extent of distortion hotspots ({\em i.e.}, their length along the mean limit cycle),
we analytically estimate the typical time interval for a stochastic trajectory  to return to 
the mean limit cycle, assuming that it has already significantly deviated from it.  Assuming a relatively constant angular velocity 
about the limit cycle, one can then estimate the limit cycle arclength required for the particle to return to the mean limit cycle after such a noise-induced deviation. In this way, we obtain a rough 
measure of the size of the regions of the limit cycle where one can expect significant noise-induced distortions.  Identifying points on the limit cycle where trajectories 
are likely to diverge from the mean limit cycle and estimating the typical extent of distortion hotspots allows one to predict from the underlying deterministic 
equations which parts of the limit cycle are inherently more susceptible to noise.

To address this question, we consider a trajectory that starts at some fixed distance from the mean limit cycle.  We choose this distance using the criterion that the system's deviation has 
increased its potential energy to $3 T$ above the minimum (which occurs at or near the mean limit cycle in the limit of a weak drive).  We treat the stochastic dynamics of the system in the plane perpendicular to the limit cycle, which we assume here to be one dimensional (higher dimensional generalizations are possible). For the analytic estimate, we consider the confining potential to be locally quadratic, an approximation warranted by the measured confining potential plotted along the local normal to the limit cycle in Fig.~\ref{fig:Potential-energy-along-limit-cycle}. We do not include a local nonzero normal component of the drive force, but the calculation can be readily generalized to include a roughly constant force term.

Using these simplifications, we compute the mean first passage time distribution for the system to return to the potential minimum.  The details of the calculation are presented in Appendix~\ref{app:first-quad}. $\kappa$ denotes the curvature of the confining potential, and its variation around the limit cycle is illustrated in Fig. \ref{fig:kappa-plot}. As explained in Appendix~\ref{app:first-quad}, we
compute the integrated survival probability $N(t)$ of trajectories starting at a fixed normal distance from the mean limit cycle and vanishing upon their return to it. The negative time derivative of this quantity is the probability distribution of the first return time.  We plot the integral $N(t)$ as it is less susceptible to noise in the numerical data.  Given a starting position $x_{0}$ in a 
harmonic potential with curvature $\kappa$, we find the integrated survival probability to be
\begin{equation}
N(t) = {\rm erf}\left[ \left(\frac{\kappa}{2k_{\rm B}T(1-e^{-2 t \kappa  B})}\right)^{1/2} x_0 e^{-t\kappa B} \right].
\label{eq:quad-soln}
\end{equation}
Here B is the mobility of the overdamped system (which is set to 1 in our simulations, without loss of generality) and $k_{\rm B}T$ is a measure of the amplitude of the Gaussian white noise.  In our simulations, $k_{\rm B}$ is normalized to 1. 

We plot $\kappa$, the curvature of the confining potential in the direction normal to the limit cycle, of our generalized Hopf model as a color map superposed on the limit cycle in Fig~\ref{fig:kappa-plot}.  As expected, corner-cutting occurs where that potential is smaller than average.  More significantly, we plot the decay of
$N(t)$ predicted solely from that local curvature, for two representative parts of the limit cycle:  (1) a region of small $\kappa$ (upper left), where the distribution of the return times is 
broad, indicating that many trajectories deviate from the mean path over significant portions of the limit cycle, and (2) a region of high curvature (lower left), where trajectories that do deviate rapidly return to the mean path.  
\begin{figure}
	\includegraphics[width=1\linewidth]{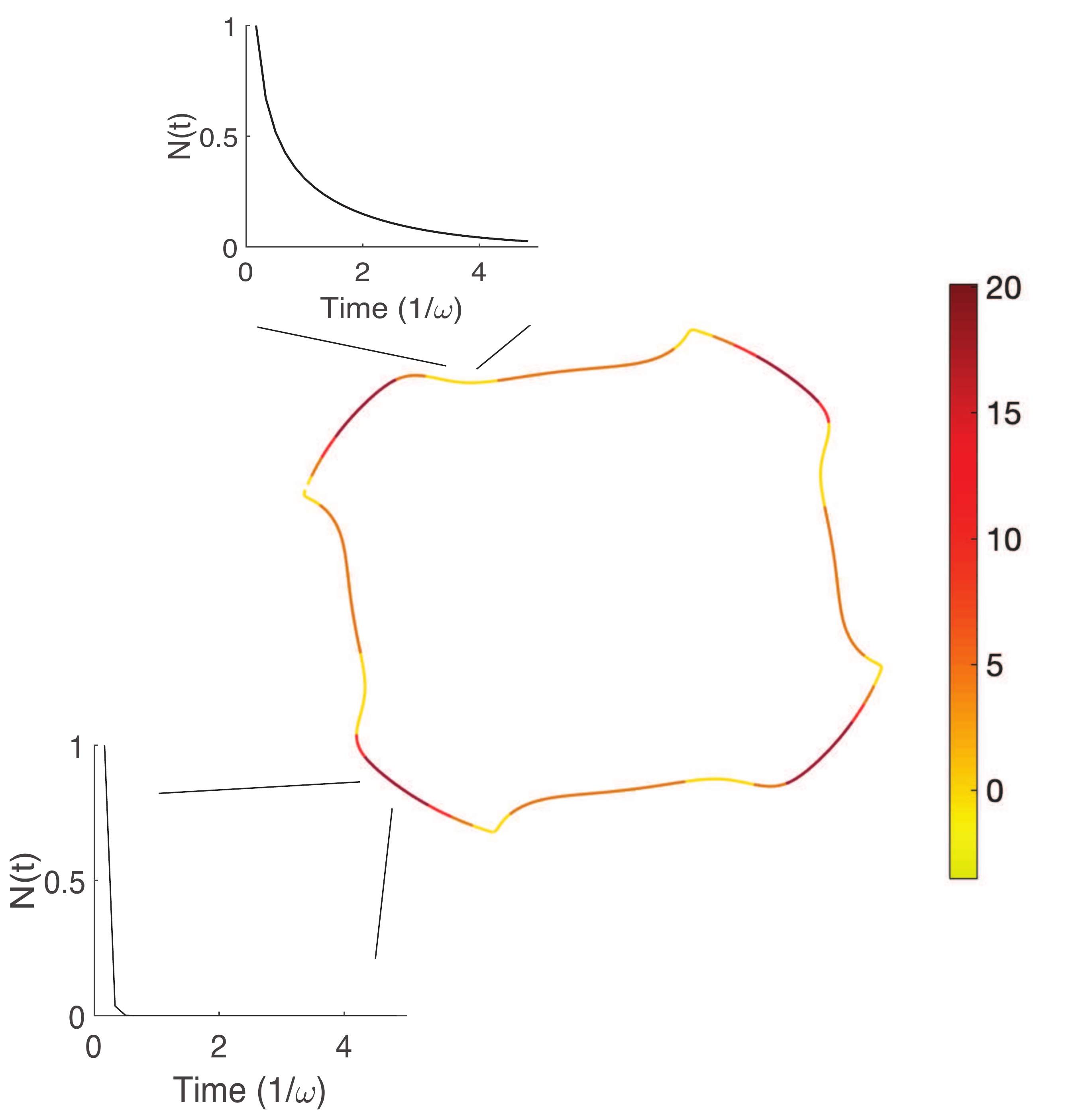}
	\colorcaption{{\bf Curvature of the confining potential:} The plot depicts $\kappa$ values along the zero-temperature limit cycle. We depict the inner curve, since the asymmetry of the problem, renders it more susceptible to corner-cutting. Additional plots exhibiting Eq.~\ref{erf}, illustrate the time of decay for the total number of trajectories that have escaped the mean path of a system with noise variance $\langle \eta^2 \rangle = 30$. This points to a theoretical method of determining regions in the oscillatory system that are prone to distortion in the presence of noise, and hence less reliable when fitting parameters to experimental data.
		\label{fig:kappa-plot}}
\end{figure}

\begin{figure}
	\includegraphics[width=1\linewidth]{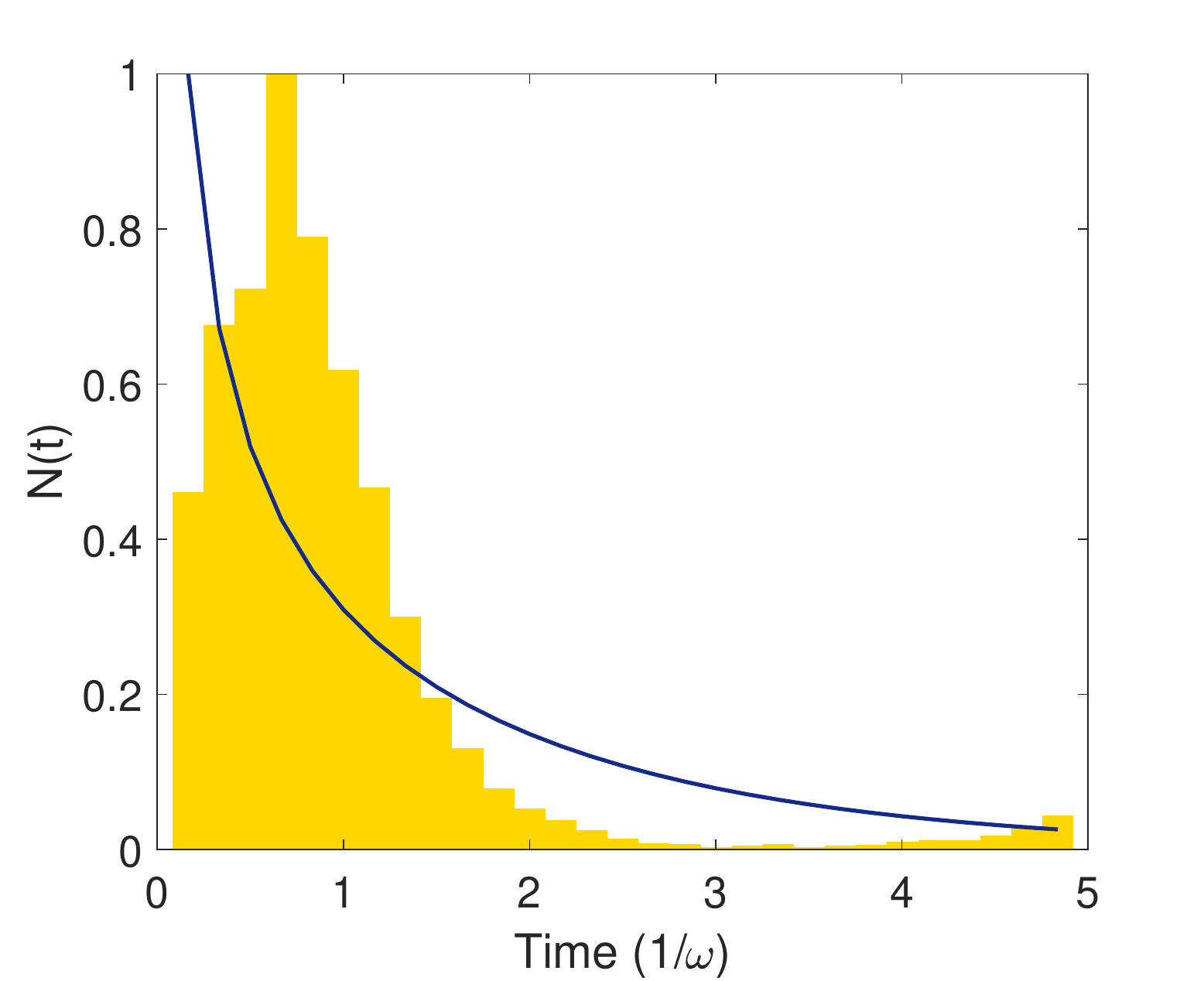}
	\colorcaption{{\bf Return time distribution:} The histogram represents the distribution of the stochastic trajectories that lie $3 T$ above the minimum potential. We consider
	trajectories that leave the mean limit cycle in a region of small potential curvature ($\kappa = 0.6$) as shown in Fig.~\ref{fig:Potential-energy-along-limit-cycle}B. The overlaying plot is the 		theoretical prediction of Eq.~\ref{eq:quad-soln}.
		\label{fig:comparing-theory}} 
\end{figure}

To test this analytic prediction for the return time distribution, we use stochastic numerical simulations to compute the distribution of return times for trajectories that start off the mean path, using the 
criterion discussed above.  The histogram of those return times is plotted (yellow bars) in Fig.~\ref{fig:comparing-theory}. The numerical data are taken from a region where the 
confining potential is weak so that such large excursions from the mean are relatively common, allowing us to obtain a larger data set of deviant trajectories.
Superimposed on this plot is the integrated survival probability 
$N(t)$ computed from Eq.~\ref{eq:quad-soln}.  There are no free fitting parameters. 

We observe reasonably good agreement between the simple model and the numerical data.  The largest discrepancies appear to be that the simple model overestimates the rapid return times and  underestimates the return times that are of the order of $\sim 1/\omega$.  We believe that this error results from our neglect of the vector potential force, which changes rather rapidly in this 
portion of the limit cycle.  As shown in Fig~\ref{fig:fv-30T}, trajectories leaving the limit cycle at this point experience initially the normal component of the driving force, and this 
normal component of the force decays rapidly as the system traverses the limit cycle.  The result is that rapid returns are suppressed by the vector potential force, but this suppression of 
returns vanishes quickly as the particle continues on its trajectory.  While the details are not captured by this simple quantitative estimate provided by Eq.~\ref{eq:quad-soln}, it has qualitative value in 
predicting regions of the limit cycle where large noise-induced deformations are likely to occur. We note (data not shown) that in regions of large $\kappa$, we observe few large excursions from the 
limit cycle and rapid returns when such excursions do occur.  

\section {\label{sec:summary}Summary}

We have shown that the fluctuations of a stochastic nonlinear oscillator can affect the size and distort the shape of its average limit cycle, as a function of noise amplitude. This effect appears 
to be dominated by particular parts of the limit cycle that combine two special features. First, the confining 
potential that stabilizes the deterministic limit cycle is broad, and second, the non-conservative driving force has a significant component normal to the local tangent of the
limit cycle. These two criteria provide a way to determine quantitatively how susceptible the deterministic limit cycle is to noise-induced distortions. Since the criteria for large deformation occur near high curvature parts of the deterministic limit cycle, we refer to these distortions as {\em corner-cutting} events.   We also provide a simple 
estimate of the duration of large, noise-induced excursions from the typical path of the nonlinear system and thereby provide a measure for the size of the noise-deformed regions of the 
limit cycle. 

Using that estimate, one is able to predict which features of the limit cycle of a periodic, nonlinear, dynamical system  are susceptible to noise and which are not.  This leads to two observations.
First, we believe that, in using noisy experimental data to fit parameters of complex nonlinear models, one must first determine which parts of the limit cycle of the dynamical system are least susceptible to that noise and weight the fits of the various model parameters accordingly. This is particularly true in systems where the details of the noise sources are poorly understood and, as a consequence, the expected noise amplitude is unknown.  Secondly, we predict that more noisy dynamical systems will generically have fewer high curvature features in their limit cycles due to corner cutting.  This trend has not yet been confirmed to our knowledge. 

Applying these findings to models of biological systems in general, and hair cells in particular, we suggest that increasing the complexity of dynamical models provides diminishing returns: more sophisticated models typically introduce new and finer features to their limit cycles, which we show will be smoothed by averaging over stochastic trajectories. For active systems
exhibiting a limit cycle, increasing amount of averaging brings one arbitrarily closer to the mean limit cycle, not to the deterministic one. The effect is analogous to that of the thermal expansion of
crystals where a combination of thermal noise and a nonsymmetric potential lead to temperature-induced changes to the mean atomic spacing.  Hence, if the presence of realistic noise amplitudes in the model leads to a significant distortion of the mean limit cycle, any finer features of the deterministic model will be inherently inaccessible to experiment.

Since we expect noisy limit cycle oscillators to not typically exhibit sharp features in their limit cycles regardless of the complexity of their underlying dynamical models, 
one may wish to investigate them more closely in noisy biological systems.  Their presence should be atypical at least, and such features 
imply tight dynamical control through very 
large curvatures of the effective confining potential.  That tight control may point to selection pressure on the relevant dynamical features of the biological limit cycle, although 
other interpretations would remain possible.

There are a number of extensions to this analysis that can be considered. First, one may examine the role of colored (frequency-dependent) noise in the system.  Here, we expect that 
increasing the noise amplitude at low frequencies will produce larger scale distortions than those at high frequencies.  The quantitative details of this effect have not been pursued yet.  Further, one may consider more complex issues, such as stochastic variations in the model parameters themselves.  These will generally introduce multiplicative noise in the system and 
render the problem significantly more complex.  We expect, however, that basic features explored here will still provide a rough set of guidelines for determining what parts of
the limit cycle are susceptible to internal stochastic forces.

\acknowledgements

DB acknowledges support from NSF PoLS grant 1705139. AJL acknowledges partial support from NSF-DMR-1709785.

	\newpage

\appendix

\section{Simulation details}
The stochastic simulations of Eq.~\ref{Hopf-main} were carried out using the 4$^{th}$-order Runge-Kutta 
method for a duration of 60 s, which corresponds to approximately 6500 limit cycles. The time  
steps used in the simulation were in the range of $10^{-4}  \leftrightarrow 2 \times 10^{-3}$ s.  The time steps for the simulations of 
Eqs.~\ref{scalar-general-hopf},\ref{vector-general-hopf} were $6 \times 10^{-7} \leftrightarrow 3 \times 10^{-6}$.
We did not observe any numerical instabilities of the solution during these runs. 

We explored a large range in the amplitude of the noise variance $\langle \eta_Z^2 \rangle$, covering the range of $10^{-7} \leftrightarrow  0.4$ where the amplitude of the 
limit cycle oscillator oscillation amplitude was held to be ${\cal O}(1)$.  The stochastic terms driving the dynamical variables  $\left\{ X(t), Y(t) \right\}$ were always 
assumed to be uncorrelated.

\section{First passage time distribution for a quadratic confining potential}
\label{app:first-quad}
To estimate the distribution of return times over which corner cutting trajectories come back to the mean limit cycle, we consider 
a simple Smoluchowski equation giving the time evolution of the probability distribution of the normal distance of a trajectory from the mean
limit cycle.  We make a number of simplifying assumptions.  First, we assume that the effective potential for this one dimensional problem is fixed in time. 
In the actual system, this potential is time varying as the particle traverses it trajectory, but as long as the excursions from the mean limit cycle are sufficiently brief, 
this approximation should provide a reasonable estimate of the return probabilities. Secondly, we assume that the force associated with the vector potential 
may be ignored.  We find that this non-potential force is typically subdominant; in principle, a time-independent approximation to this force could be included in the 
analysis explored below by adding a constant force, corresponding to a simple tilt of the potential landscape.  Finally, the landscape of that confining potential is 
assumed to be locally quadratic, as illustrated by panels (C) and (D) in Fig.~\ref{fig:Potential-energy-along-limit-cycle}.

Given these approximations, we may write the Smoluchowski equation as
\begin{equation}
\label{fokker-plank}
\frac{\partial P(x,t)}{\partial t} =D \frac{\partial^2 P(x,t)}{\partial x^2} + \kappa B \frac{\partial xP(x,t)}{\partial x}
\end{equation}
where,
$D =B k_{\rm B} T$ is the effective diffusion constant and $B$ the mobility.  $\kappa$ is curvature of the 
confining potential, which may be computed directly from the equations of motion and the curve associated with the mean limit cycle.   Using this
equation we will compute the probability that a trajectory, starting at a particular normal distance from the mean limit cycle, returns to that mean 
limit cycle for the first time after a time interval $t$.  This is the well-known first passage time distribution. 

We note that Eq.~\ref{fokker-plank} has a simple time-independent solution corresponding to the equilibrium position distribution of a harmonic oscillator with 
spring constant $\kappa$:
\begin{equation}
\label{stationary-solution}
P_{\rm st}(x,t) = \sqrt{\frac{\kappa}{2\pi k_{\rm B}T}}e^{-\frac{\kappa x^2}{2 k_{\rm B} T}}
\end{equation}
Writing the time-dependent probability distribution that evolves towards $P_{\rm st}(x)$ according to Eq.~\ref{fokker-plank} as a product: 
$P(x,t) = P_{st}(x,t)^{1/2} g(x,t)$, we obtain a new evolution equation for $g(x,t)$:
\begin{equation}
\label{equation-g(x,t)}
\frac{\partial g(x,t)}{\partial t} -B k_{\rm B} T\frac{\partial^2 g(x,t)}{\partial x^2} + \frac{B\kappa}{2}(\frac{\kappa}{2 k_{\rm B} T} - 1)g(x,t) = 0.
\end{equation}
We note that the $g \longrightarrow 1$ at long times in order to be consistent with Eq.~\ref{stationary-solution}.  

Using separation of variables, $g(x,t) = f(t)h(x)$ and simple redefinition of the curvature $\frac{\kappa}{2 k_{\rm B} T} = \beta$, we find that $f$ and $h$ obey the 
ordinary differential equations:
\begin{eqnarray}
\label{equation-f(t)}
\frac{df}{dt}  + \frac{f}{\tau}&=& 0 \\
\frac{d^2 h}{dx^2} - \beta^2 x^2 h + (\beta + \frac{1}{\tau D}) h &=& 0.
\label{equation-h(x)}
\end{eqnarray}
From Eq.~\ref{equation-f(t)} we see that $g(x,t)$ decays exponentially in time with decay rates $\tau^{-1}$ set by solutions of the Eq.~\ref{equation-h(x)}.  That equation 
may be reduced Hermite's differential equation via a rescaling of both the independent $y = \sqrt{\beta}x$ and dependent $h(y) = u(y)e^{-\frac{y^2}{2}}$ variables: 
\begin{equation}
\label{equation-u(y)}
\frac{d^2 u}{dy^2} - 2y\frac{du}{dy} + \frac{u}{\beta \tau D} = 0
\end{equation}

The eigenfunctions $H_n(y)$ of this differential operator 
\begin{equation}
\label{hermite-eigenfunctions}
H_n(y) = (-)^ne^{y^2}\frac{\partial^n e^{-y^2}}{\partial y^n}
\end{equation}
allow us to determine the discrete set of decay rates
\begin{equation}
\label{hermite-eigenvalues}
\tau_n^{-1} = 2n\beta D.
\end{equation}

Combining Eqs.~\ref{stationary-solution},\ref{equation-f(t)},\ref{equation-h(x)} and,\ref{hermite-eigenfunctions}, we write the solution to Eq.~\ref{fokker-plank} (in terms of the scaled 
spatial variable $y$) as 
\begin{equation}
P(y,t) = e^{-y^2}\sum_{n} c_n H_n(y)e^{-\frac{t}{\tau}},
\end{equation}	
where the undetermined coefficients $c_n$ are given by the initial condition: $P(y,t=0)$. We take that initial condition to be a delta function $\delta(x-x_0)$, where $x$ measures the 
normal displacement from the the mean limit cycle and $x_{0}$ is set by choosing the point where the potential energy of the system is $3 k_{\rm B} T$ above that of the mean limit cycle.
From the orthonormality of the Hermite polynomials,
\begin{equation}
\label{orthogonality-hermite}
\int_{-\infty}^{\infty} dy e^{-y^2}H_n(y)H_m(y) = \delta_{mn}2^m m! \sqrt{\pi},
\end{equation}
we obtain the undetermined constants in terms of $ y_{0} \sqrt{\beta} x_{0}$: 
\begin{equation}
\label{equation-cn}
c_n = \frac{H_n(y_0)}{2^n n! \sqrt{\pi}}
\end{equation}

From these we have the conditional probability
\begin{equation}
\label{probability-y}
P(y,t|y_0, t_0) = \sqrt{\frac{\kappa}{2\pi  k_{\rm B} T}} e^{-y^2} \sum_{n} \frac{H_n(y) H_n(y_0)}{n!} \left(\frac{e^{-{t\kappa B}}}{2}\right)^n
\end{equation}
that a trajectory starting at $y_{0}$ at time zero reaches $y$ at time $t$.   Returning to the unscaled independent variable and using Mehler's approximation we write
\begin{equation}
\label{probability-x}
P(x,t|x_0,t_0) = \left(\frac{\kappa}{2\pi  k_{\rm B} T(1 - e^{-2\kappa B t})}\right)^{1/2} e^{-\frac{\kappa(x - x_0 e^{-t \kappa B})^2}{2 k_{\rm B} T(1-e^{-2\kappa Bt})}}.
\end{equation}

In order to ensure we compute the first passage time to the origin, we must eliminate trajectories that pass through $x=0$ on their way to $(x,t)$.  We do so in the usual way by 
introducing an absorbing boundary condition at the origin.  This is simply accomplished by subtracting the above result from an imagined solution: 
\begin{equation}
\tilde{P}(x,t) =  P(x,t) - P(-x,t).
\end{equation}
Using this result, we 
compute the total probability remaining at time $t$: 
\begin{equation}
N(t) = \int_{0}^{\infty} \tilde{P}(x,t)dx.
\end{equation}

The resulting integral can be written as
\begin{equation}
\label{N(t)-x}
\begin{split}
N(t) = (\frac{\kappa}{2\pi  k_{\rm B} T(1 - e^{-2\kappa B t})})^{1/2} \times &\\
\int_{0}^{\infty}dx \left\{  e^{-\frac{\kappa(x - x_0 e^{-t \kappa B})^2}{2 k_{\rm B} T(1-e^{-2\kappa Bt})}} \right.
 & -  \left. e^{-\frac{\kappa(x + x_0 e^{-t \kappa B})^2}{2 k_{\rm B} T(1-e^{-2\kappa Bt})}} \right\}
\end{split}
\end{equation}
The remaining integral is easily performed to yield a solution written in terms of the error function:
\begin{eqnarray}
N(t) &=& \frac{2}{\sqrt{\pi}} \int_{0}^{y_0} e^{-(y-y_0)^2}dy\\
&=& {\rm erf}\left[ \left(\frac{\kappa}{2 k_{\rm B} T(1-e^{-2\kappa Bt})}\right)^{1/2} x_0 e^{-t\kappa B} \right].
\label{erf}
\end{eqnarray}
This result appears in the main text -- see Eq.~\ref{eq:quad-soln}.

\end{document}